# Gravitational-wave population inference with deep flow-based generative network

Kaze W. K. Wong,[1,*] Gabriella Contardo,[2] and Shirley Ho[2,3,4]

[1]*Department of Physics and Astronomy, Johns Hopkins University, 3400 N. Charles Street, Baltimore, Maryland 21218, USA*
[2]*Center for Computational Astrophysics, Flatiron Institute, New York, New York 10010, USA*
[3]*Department of Astrophysical Sciences, Princeton University, Princeton, New Jersey 08540, USA*
[4]*Physics Department, Carnegie Mellon University, Pittsburgh, Pennsylvania 15213, USA*

We combine hierarchical Bayesian modeling with a flow-based deep generative network, in order to demonstrate that one can efficiently constraint numerical gravitational-wave (GW) population models at a previously intractable complexity. Existing techniques for comparing data to simulation, such as discrete model selection and Gaussian process regression, can only be applied efficiently to moderate-dimension data. This limits the number of observable (e.g., chirp mass, spins.) and hyperparameters (e.g., common envelope efficiency) one can use in a population inference. In this study, we train a network to emulate a phenomenological model with 6 observables and 4 hyper-parameters, use it to infer the properties of a simulated catalogue and compare the results to using a phenomenological model. We find that a 10-layer network can emulate the phenomenological model accurately and efficiently. Our machine enables simulation-based GW population inferences to take on data at a new complexity level.

## I. INTRODUCTION

Since the discovery of gravitational waves (GWs) [1], GW events are being detected routinely at an accelerating pace. By the end of the third observational run (O3) of the ground-based GW detector network, which will end on 31 April 2020, one can anticipate ∼60 detected events in total. The ground-based GW detector network, including the Laser Interferometer Gravitational-Wave Observatory (LIGO), the Virgo interferometer [2], and the Kamioka Gravitational Wave Detector (KAGRA) [3] (henceforth *LVK collaboration*), is expected to operate at its design sensitivity in late 2021, which will detect ∼100 events per year [4]. And with the planned A+ upgrade, we will detect hundreds to a thousand events per year [5]. With the growing GW catalogue, the focus of gravitational wave astrophysics will rapidly shift toward studying the population of GW events [6,7]. The population of GW events offers a unique window into a plethora of physics, including fundamental physics such as modifications to general relativity (GR) [8], the expansion rate of the universe [9], and astrophysics related to the progenitor of the binaries [10–13]. The growing catalogue of stellar-mass compact binary systems detected by GWs offers burgeoning insights to the physics governing their evolution. At the same time, its increasing complexity demands more sophisticated modeling and data analysis techniques.

Current state-of-the-art GW analyses employ phenomenological parametric models to describe the GW population [14–18]. This is advantageous for its simplicity and agnostic to physical assumptions. On the other hand, it cannot provide much physical insight directly precisely because of the same reason. Alternatively, there are simulations which create synthetic populations of GW events based on some physical assumptions which are characterized by a set of parameters, such as the metallicity of the environment [19] or the escape speed of the stellar cluster in which the GW-emitting binary resides [20]. One can in principle compare the simulation results to the data and obtain direct constraints on these physical parameters. This is often done by calculating the Bayes factor between models with different parameters under the same parametrization [21–23]. In practice, the simulation-based approach has an obvious disadvantage—simulations are often computationally heavy. Obtaining constraints on the physical parameters requires good sampling in the parameter space of interest, which in turn would require a simulation on each sample point, and the heavy computational load of each simulation basically renders this thorough sampling impractical.

There are recent developments in circumventing this technical difficulty by creating an emulator of the simulations with machine learning techniques [24,25]. They emulate the output of the simulations with Gaussian process regression (GPR) and principal component analysis (PCA) without going through the sophisticated simulations, hence gaining enough speed up so that the emulator can be used in the population inference. Despite the novelty demonstrated in previous studies, this existing method has a few limitations:

(i) Data with high complexity, such as simulations parametrized by a large number of hyperparameters, will require a decent number of training simulations to reach a given accuracy. Since the machine is running Gaussian processes on every principal component (PC) that characterize the entire set of simulations, which the number scales roughly linearly with the number of simulations, the GPR-PCA machine becomes progressively inefficient in emulating simulations as the complexity of the set of simulations increases.[1] Also, PCA is used to compress the dimension of the training data in order to reduce the computational load, which means there is an inevitable loss of information.

(ii) As mentioned in the first point, the computational load of a GPR-PCA machine scales linearly with the number of PCs after the compression. If one wants to maintain the same speed of the machine, more PCs need to be discarded, which degrades the quality of the interpolation and potentially introduces bias into the final result. Another issue is that the existing machines use histograms to estimate the probability density function of their observable space, which means their estimate of the probability density is binning-dependent.

(iii) Moreover, the memory requirement for a histogram increases geometrically with the dimension of the problem. For example, we want to have 6 observables in our problem, and we choose to bin our observables into 30 bins in each observable axis. Assuming we use FLOAT32 to store the histograms which we use to train the machine, and we have 1000 simulations, then the size of the training data to be loaded in memory will be ∼2.5 TB, which is not available on most computer cluster. Therefore the existing machine cannot be easily generalized to higher-dimensional problems.

---

[*]kazewong@jhu.edu

In this work, we demonstrate the technique of normalizing flows (a type of flow-based generative model) can overcome these difficulties, hence improving the efficiency and accuracy of simulation-based GW population inference. More specifically, we use a normalizing flow network to emulate the likelihood in a hierarchical Bayesian analysis (HBA) framework. The field of deep learning is growing exponentially due to advancements in various fields: innovations on the topic of neural networks [26,27], increasingly powerful hardware [28,29], development of open source general purpose deep learning libraries [30,31], etc. These advancements lower the barrier to apply deep learning techniques to problems in other fields. In particular, the astronomy and astrophysics community has applied deep learning techniques in various aspects, including signal detection [32–37], inference [38–41], and simulations [42]. Deep learning techniques often offer more flexible and much faster solutions compared to traditional methods. Normalizing-flows models rely on using a series of simple invertible transformations to map a complicated distribution (usually the data observed) to a simple one (e.g., a multivariate Gaussian). This specific formulation can provide an estimate of the log-probability for a given datapoint in the original distribution. They can therefore be used for density estimation, see for instance [43].

This paper is structured as follow: In Sec. II, we describe the specific problem of GW population inference and review the HBA framework. In Sec. III, we review the basics of normalizing flows, layout the architecture of our network. In Sec. IV, we present our data and results. In Sec. V, we discuss prospects of this work.

## II. HIERARCHICAL BAYESIAN INFERENCE

In this section, we summarize the salient points of hierarchical Bayesian inference and clarify our objective. We refer interested readers to more detailed explanations in the literature [44–46]. GW data is usually given as a time series with some characteristic waveform. In order to extract physical quantities from the time series, such as masses and spins of a GW-emitting binary, one often adopts a parameter estimation process with Bayesian inference [47]. Given some data $d$, the posterior probability of the signal being an astrophysical source with parameters $\theta$ is given by $p(\theta|d) \propto p(d|\theta)p(\theta)$, where $p(d|\theta)$ is the likelihood of observing the data given our model of the astrophysical signal and detector, and $p(\theta)$ is the prior we assume on the source parameters. The prior encodes our intuition on the underlying physics (for example, mass should not be negative), and plays an important role in interpreting the result [48,49].

A hierarchical analysis parameterizes the prior used in parameter estimation of a GW event with some *hyperparameters* $\lambda$, so that we can infer the true hyperparameters by marginalizing over the event parameters,

$$p(\lambda|d) = \frac{p(\lambda) \int p(d|\theta)p(\theta|\lambda)d\theta}{p(d)}. \qquad (1)$$

The term $p(d|\theta)$ is the single event likelihood, whereas $p(\lambda)$ is now a prior on the hyperparameter. $p(\theta|\lambda)$ is the *population likelihood*. For clarity, we denote the parameters which describe the individual GW event properties as *event*

---

[1]The more complex the simulations are, the harder it is to compress the entire set of simulations to the same number of principal components.

*parameters*, $\boldsymbol{\theta}$, and the parameters which describe the entire set of events as *population parameters*, $\boldsymbol{\lambda}$. As an example, in studies which take the route of employing phenomenological models, $p(\boldsymbol{\theta}|\boldsymbol{\lambda})$ could be a power law in mass, where the spectral index $\alpha$ is the population parameter and the mass $m$ is the event parameter, i.e., $p(m|\alpha) \sim m^\alpha$. In contrast, a simulation-based model often provides a synthetic catalogue of GW event, instead of an analytical expression of the population likelihood. The objective of this study is to find an efficient way to construct an emulator of the population likelihood $p(\boldsymbol{\theta}|\boldsymbol{\lambda})$ given a set of GW catalogue generated by numerical simulations, so we can then use the emulator in Eq. (1) to compute the population posterior.

For a set of events which are drawn independently from the same underlying population, and if the parameter estimation of different astrophysical events is not correlated (i.e., the signals are not overlapping), the likelihood of observing that particular set of events can be factorized into the product of the individual event likelihoods,

$$p(\boldsymbol{d}|\boldsymbol{\lambda}) = \int p(\boldsymbol{d}|\{\boldsymbol{\theta}\}) p(\{\boldsymbol{\theta}\}|\boldsymbol{\lambda}) d\{\boldsymbol{\theta}\} \qquad (2)$$

$$= \prod_{i=1}^{N_{\text{obs}}} \int p(\boldsymbol{d}_i|\boldsymbol{\theta}_i) p(\boldsymbol{\theta}_i|\boldsymbol{\lambda}) d\boldsymbol{\theta}_i. \qquad (3)$$

Note that $\boldsymbol{d}$ in Eq. (2) is the entire time series observed by the GW detector network, while $\boldsymbol{d}_i$ is Eq. (3) refers to the segment which contains the event characterized by $\boldsymbol{\theta}_i$. The term $p(\boldsymbol{d}_i|\boldsymbol{\theta}_i)$ is usually rewritten as $p(\boldsymbol{\theta}_i|\boldsymbol{d}) p(\boldsymbol{d}_i)/p(\boldsymbol{\theta}_i)$ using Bayes' theorem. Combining Eq. (1) and Eq. (3), we obtain the population posterior,

$$p(\boldsymbol{\lambda}|\boldsymbol{d}) = p(\boldsymbol{\lambda}) \prod_{i=1}^{N_{\text{obs}}} \int \frac{p(\boldsymbol{\theta}_i|\boldsymbol{d}) p(\boldsymbol{\theta}_i|\boldsymbol{\lambda})}{p(\boldsymbol{\theta}_i)} d\boldsymbol{\theta}_i. \qquad (4)$$

The event posterior PDF $p(\boldsymbol{\theta}_i|\boldsymbol{d})$ is often given in the form of $S$ discrete samples by a parameter estimation process [50,51]. We can now make use of the posterior samples produced by a separate parameter estimation pipeline, thus avoiding unnecessary re-computation of estimating $p(\boldsymbol{d}|\boldsymbol{\theta})$ and reducing the computation load for each population inference run significantly. The integral in Eq. (4) is essentially the expectation value of the prior-reweighted population likelihood, which can be turned into a discrete sum over the event posterior PDF samples:

$$p(\boldsymbol{\lambda}|\boldsymbol{d}) = p(\boldsymbol{\lambda}) \prod_{i=1}^{N_{\text{obs}}} \frac{1}{S_i} \sum_{j=1}^{S_i} \frac{p(^j\boldsymbol{\theta}_i|\boldsymbol{\lambda})}{p(^j\boldsymbol{\theta}_i)}, \qquad (5)$$

where $j$ labels the $j$th sample of the $i$th event. Our models in this work do not predict the event rate, so we also leave the rate out in deriving Eq. (5). Since the event rate is integrated over all the event's parameters, it is solely a function of the hyperparameters, so traditional interpolation methods can handle the rate, and it can be incorporated into the inference machine trivially.

Parameter estimation of GW events comes with its own systematics [52–54] and the computation is often quite time consuming. In the limit of high signal-to-noise ratio (SNR), measurement uncertainties are negligible and the inferred parameters of an event will be distributed as a Gaussian around the true value [55], with standard deviation inversely proportional to the SNR. To avoid complication and unnecessary use of computational resources, in this study we take the high-SNR limit and treat the measured events as if they had no measurement systematics and statistical uncertainties on the event parameters, i.e., $p(\boldsymbol{\theta}_i|\boldsymbol{d}) = \delta(\boldsymbol{\theta}_{i,\text{true}} - \boldsymbol{\theta}_i)$. Then the integration in Eq. (4) will simply pick out the correct value, therefore the posterior can be written as

$$p(\boldsymbol{\lambda}|\boldsymbol{d}) = p(\boldsymbol{\lambda}) \prod_{i=1}^{N_{\text{obs}}} p(\boldsymbol{\theta}_i|\boldsymbol{\lambda}). \qquad (6)$$

In the high-SNR limit, the prior on event parameters is irrelevant [55] and can be treated as a constant. Uncertainties in the selection biases will propagate through the analysis pipeline, resulting in an additional systematic error in the result when analysing real data [56,57]. In this paper, our goal is to demonstrate that this deep learning technique can enable the use of more sophisticated simulations in population inference, and we are generating simulated data for this purpose. Leaving the selection biases out avoids the confusion between systematic from the deep learning method and inaccuracy in the selection biases. We focus here on the ability of the model to recover the hyperparameters $\boldsymbol{\lambda}$ without considering the effect of selection biases, so any systematics will be due to the inaccuracy in the interpolation.

In reality, the uncertainties and systematic biases in event parameter estimation will propagate to the population inference result, smoothing out the population posterior and adding biases to the inferred population parameters. This means that systematic biases induced by the inaccuracy of our interpolation method will become less significant when we include measurement uncertainties and biases in our analysis, so the results we present in this work are conservative.

## III. NORMALIZING FLOWS

As posed in Sec. II, the central problem we are trying to tackle in this paper is: Given a set of simulated catalogues of GW events, with each event characterized by a vector of event parameters $\boldsymbol{\theta}$ and each catalogue labeled by a vector of hyperparameters $\boldsymbol{\lambda}$, can we construct a function that can approximate $p(\boldsymbol{\theta}|\boldsymbol{\lambda})$ with satisfying accuracy, and at the same time can be evaluated fast enough to be used as the population likelihood in a hierarchical Bayesian analysis?

We approach this question using a conditional neural density estimator, in particular a flow-based generative (often referred as normalizing flow) model. Flow-based generative models have been recently developed and explored in the machine-learning community [43,58–62]. This family of models has a unique perk compared to other neural generative models. In addition to being able to perform good data generation, they also provide an estimate of the probability density function of the data. This makes this family of model a perfect fit for our task. In this section, we present the general principles behind the model we employ.

Normalizing flow propose to transform a simple density (e.g., a Gaussian) $z \sim p_z$ into a target, complex, density (the data) $x \sim p_x$. Building a model to learn a mapping $g$ from $z$ to $x$ is a common idea in generative models, and several methods have been proposed, e.g., using discriminator networks (generative adversarial networks, [27]), or approximate inference (variational auto encoders [63]). However, these approaches do not allow us to evaluate $p_x$, but can only generate new data that mimics $p_x$. On the other hand, normalizing flow [58,64] propose to use a mapping function $g: \mathbb{R}^d \to \mathbb{R}^d$ that is invertible (bijective), with a tractable Jacobian. This will allow us to learn the mapping using directly the maximum likelihood, through a change of variable, with which we can compute the normalized probability density $p_x(x)$ from $p_z(z)$ when we use an invertible function:

$$p(\mathbf{x}) = p_z(\mathbf{z}) \left| \det\left(\frac{\partial g(\mathbf{z})}{\partial \mathbf{z}^T}\right) \right|^{-1}. \quad (7)$$

Equation (7) is tractable as long as $g$ is easily invertible and the determinant of its Jacobian is easy to compute, hence normalizing flow only require a careful design of $g$. Interestingly, if two functions $g_1$ and $g_2$ are both invertible and have a tractable Jacobian, the composition $g_1 \circ g_2$ also has these properties. Additionally, generating new data $x$ can be done by drawing a sample $z \sim p_z$ and computing the value of $x$ through the set of transforms in the normalizing flow network, $x = f^{-1}(z) = g(z)$. As written above, the density on a given data point $x$ can be computed as the density of its image $f(x)$ multiplied by the determinant of the corresponding Jacobian.

Instead of choosing one complicated transform which maps our prior to the target density distribution, we can restrict the network to use a series of $K$ simple transforms $g_k, k = 1, ..., K$, which are invertible and whose Jacobians can be easily calculated. We can then apply Eq. (7) repeatedly to obtain arbitrarily complex probability density distributions, given enough number of transforms. With the series of transforms, the target random variable and its probability density distribution are given by

$$\mathbf{z}_k = g_k \circ \cdots \circ g_1(\mathbf{z_0}), \quad (8)$$

$$p(\mathbf{x}) = p(\mathbf{z}_K) = p_{z_0}(\mathbf{z_0}) \prod_{k=1}^{K} \left| \det\left(\frac{\partial g_k}{\partial \mathbf{z}_k}\right) \right|^{-1}. \quad (9)$$

As an example, we can now estimate the probability density in a relatively simple distribution such as a multivariate Gaussian, then apply Eq. (9) to obtain the probability in the target distribution. Given the transformation, we can also generate new samples from the distribution by applying the transforms to a set of samples from the prior as well.

Figure 1 illustrates the essence of normalizing flow. Given the target samples, we apply multiple transforms to a one-dimensional Gaussian distribution to fit the target.[2] After two transformations, the Gaussian prior is wrapped to fit the target sample. We also included a Gaussian kernel density estimation (kde) result for comparison. While the Gaussian kde is missing the sharp edges of the target distribution, normalizing flow can capture these features.

Choosing the correct transformation is crucial to designing an efficient network for our specific problem. Because our target is $p(\boldsymbol{\theta}|\boldsymbol{\lambda})$, it is essential that our network is capable to model conditional probabilities. A specifically designed autoregressive model known as masked autoregressive flow (MAF) [43] trivially incorporates the ability to model conditional probability. Therefore, we adopt MAF as the "flow" in the flow-based generative model. Autoregressive models [66,67] expand the joint density of a set of random variables $p(x)$ as a product of one-dimensional conditionals $p(\mathbf{x}) = \prod_i p(x_i|\mathbf{x}_{1:i-1})$. Assuming a specific order of those random variables $\mathbf{x} = \{x_1...x_D\}$, the $i$th conditional probabilities only depend on the random variables appearing prior to $x_i$, i.e., $p(x_i) = p(x_i|x_{<i})$, which $x_{<i} = \{x_1...x_{i-1}\}$. From the definition of an autoregressive model, we can understand why it naturally extends to conditional probability modeling. As long as the conditional variable $\mathbf{y}$ comes before other random variables, we can write the conditional probability of observing $\mathbf{x}$ given $\mathbf{y}$ as

$$p(\mathbf{x}|\mathbf{y}) = \prod_i p(x_i|\mathbf{x}_{1:i-1}, \mathbf{y}), \quad (10)$$

which is similar to the definition of an autoregressive model. The main drawback of this approach is its sensitivity to the order of the variables $x_i$. References [43,68] show that specific autoregressive models can be interpreted as normalizing flow (i.e., they have tractable Jacobian and are invertible). Using an autoregressive model as normalizing flow allows to increase the flexibility of the model while retaining a tractable Jacobian, but also to make the model more resilient to the ordering of variables. More specifically, in the MAF framework, a conditional-MAF simply stacks MADEs (masked autoencoder for distribution estimation [69]) functions that have been made *conditionals* by adding the condition $\lambda$ as an additional input for each layer. We can train a conditional-MAF network to emulate a given set of training GWs catalogue. Once the network is trained, it can be used as the population likelihood $p(\theta|\lambda)$ in Eq. (1).

---

[2]We used planar flow described in [65] for the illustration.

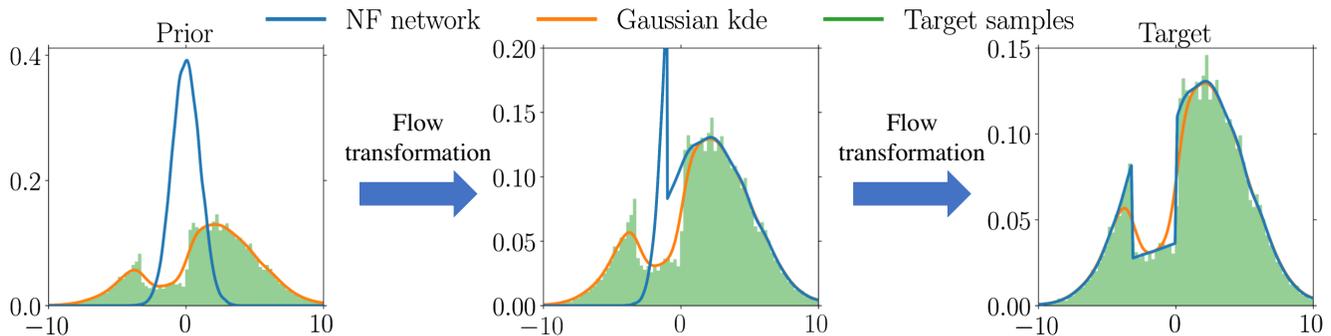

FIG. 1. An illustration of the working principle of normalizing flow. Given the target samples (histogram in green), whose pdf we want to estimate, we apply a series of transformations to wrap a Gaussian to fit the target sample. Since the probability density of the Gaussian can be obtained trivially, the pdf of the target (shown in blue) can also be computed as long as the transformations are known. In the case of a neural network, the transformation is a set of parametrized functions, whose parameters are obtained during the training of the neural network. For comparison, we also use a Gaussian kernel density estimation (kde) to estimate the pdf given the target samples. While the Gaussian kde can capture the bimodality of the target, it misses the sharp edges of the distribution. On the other hand, the normalizing flow framework reproduce the distribution with excellent accuracy.

The only difference between the model described in [43] and the model implemented in this paper are the size of input layer and network-related hyperparameters (such as the number of hidden units in a hidden layer). Therefore, we refer interested readers to Refs. [43,70] for more elaborated description of the model. We use 1024 hidden units in each layer, and we explore two variations of this model, one using 5 flow layers and the other one using 10 layers. We find the accuracy of a 10-layers network to be sufficient for our purpose.

## IV. VALIDATING AGAINST A PHENOMENOLOGICAL MODEL

To demonstrate the robustness of our deep learning interpolation method, we cross check the performance of our machine against a phenomenological model, for which we can write down an analytic expression for $p(\theta|\lambda)$, thus we can use the model directly in the population inference process and validate the results from the neural network. We generate our training data from the analytical model and train a network to interpolate the model. We use the trained network to emulate $p(\theta|\lambda)$ in a population inference analysis, then compare the inference result using the network with the result using the analytical model.

We generate our population according to the prescriptions described in [7]. The most general model family described in [7] includes 7 event parameters and 16 hyperparameters. In order to facilitate the speed of the training and inference process, we choose 6 event parameters and 4 hyperparameters as our phenomenological models. Our model includes the primary mass $m_1$, mass ratio $q$, the spin magnitude of each binary $a_i, i \in (1, 2)$ and the tilt angles $t_i$ between each BH spin and the orbital angular momentum.

For simplicity, we parametrized our mass model as two independent power laws in the primary mass and mass ratio,

$$P(m_1) \propto \begin{cases} m_1^{\alpha_m} & \text{if } m_1 \in [5, 50] \\ 0 & \text{otherwise,} \end{cases} \quad (11)$$

$$P(q) \propto q^{\beta_m} \quad q \in (0, 1], \quad (12)$$

where $\alpha_m$ and $\beta_m$ are the spectral indices of the two power laws. We choose the lower mass cutoff to be 5 $M_\odot$ and the upper mass cutoff at 50 $M_\odot$. Both are assumed to be sharp. The range of sampling in $\alpha_m$ and $\beta_m$ are chosen to be $[-3, -1]$ and $[1, 3]$, respectively. More sophisticated models [14,17,18,71] were discussed in [7], which add more free parameters into the model to capture more features in the data. However, there is no strong evidence favoring one model over another, therefore we pick a simple model in this studies.

We assume that both BH spins magnitudes are drawn from a common beta distribution [15]:

$$P(a_i|\alpha_a, \beta_a) \propto \frac{a_i^{\alpha_a-1}(1-a_i)^{\beta_a-1}}{B(\alpha_a, \beta_a)}, \quad (13)$$

where $B(\alpha_a, \beta_a)$ is the beta function. Following the choice presented in [7], we choose to model the moments of the beta distribution using the mean ($\mathbb{E}[a]$) and variance ($\text{Var}[a]$), given by

$$\mathbb{E}[a] = \frac{\alpha_a}{\alpha_a + \beta_a}, \quad (14)$$

$$\text{Var}[a] = \frac{\alpha_a \beta_a}{(\alpha_a + \beta_a)^2(\alpha_a + \beta_a + 1)}. \tag{15}$$

The pdf of a beta distribution can change quite drastically depending on the parameters characterizing the distribution, which means we will need more data and a more complex network to capture the features which represent the change in the pdf as a function of distribution parameters. For simplicity, we fix the mean and variance to be $\mathbb{E}[a] = 0.5$ and $\text{Var}[a] = 0.05$. Note that even though this choice means we do not include the change of the beta distribution as a function of the distribution parameter in the training of our model, the network still needs to fit for the beta distribution we have chosen. This resembles a practical scenario that the model one tries to interpolate has some discrete flags which affect the event parameters distribution, yet there is no need to interpolate over those discrete flags.

Finally, we follow Ref. [17] to simulate the spin orientation. We assume that the tilt angles between each BH spin and the orbital angular momentum are drawn from a mixture of two distributions: an isotropic component and a preferentially aligned component, represented by a truncated Gaussian distribution in $\cos t_i$ peaked at $\cos t_i = 1$

$$p(\cos t_1, \cos t_2 | \sigma_1, \sigma_2, \xi) = \frac{(1-\xi)}{4} \tag{16}$$

$$+ \frac{2\xi}{\pi} \prod_{i \in 1,2} \frac{e^{-(1-\cos t_i)^2/2\sigma_i^2}}{\sigma_i \text{erf}(\sqrt{2}/\sigma_i)}. \tag{17}$$

The distribution is parametrized by three parameters, $\sigma_1, \sigma_2, \xi$, which are the standard deviation of the two Gaussian and the mixing fraction between the two components. A value of $\xi = 1$ implies that all the BBH spins are preferentially aligned with the orbital angular momentum, while $\xi = 0$ implies that the spin orientations are distributed isotropically. The two components represents the two most prominent formation channels of BBH mergers: isolated and cluster formation. We fix the mixing fraction to be $\xi = 0.5$, and include $\sigma_1$ and $\sigma_2$ in our training. The range of $\sigma_1$ and $\sigma_2$ are both [0, 2]. The parameters and hyperparameters used in our phenomenological models are summarized in Table I.

Given the analytical model, we train and evaluate the performance of our machine as described below. We create the training set by sampling 100 points in the hyperparameter space with Latin hypercube sampling [72]. For each point in the hyperparameter space, we create a catalogue of $10^5$ BBH events, each characterized by the 6 event parameters, following the distribution parametrized by the 4 hyper-parameters. This means the entire training set contains $10^7$ training samples. We also create a smaller validation set, which follows the same method as creating the training set but with only 10 points in the hyperparameter space. Note that the locations of the points in hyperparameter space in the validation set is different from the training dataset. We take the state of the network for which the validation loss is minimum as our best-trained model, and use it in the inference process. The entire process of training and validating the 10 layers model takes $\sim 10$ hours on a Tesla k80 gpu.

TABLE I. Event parameters and hyperparameters used in this work.

| Event parameters $\theta$ | |
|---|---|
| $m_1$ | Primary mass in the binary |
| $q$ | Mass ratio of the binary |
| $a_1, a_2$ | Spin magnitudes of the binary |
| $\cos t_1, \cos t_2$ | Tilt angles between each BH spin and the orbital angular momentum. |
| Hyperparameters $\lambda$ | |
| $\alpha_m$ | Spectral index of $m_1$ |
| $\beta_m$ | Spectral index of $q$ |
| $\sigma_1, \sigma_2$ | Width of the preferentially aligned component of the BH spin orientation |

Figure 2 shows the interpolation result of $p(\theta|\lambda)$ for a specific $\lambda = (-2, 2, 0.5, 0.7)$, which is a test point in the hyperparameter which we have not included in both the training set and validation set. Despite the small scale difference, mainly originating from individual sampling fluctuations, both 5-layer and 10-layer models fit the large-scale behavior of the distribution quite well. In particular, the 10-layer model is performing better than the 5-layer model at the edge of $\cos t_1$ and $\cos t_2$. This is expected since a model with more layers is applying more transforms to the prior distribution, which means it is more flexible in terms of modeling a target distribution, therefore it should be able to capture more features such as the edge of a distribution.

Next, we inspect the performance of the network on a population level. The population posterior produced by a well trained network should be similar to the posterior produced by the analytical model. To test this, we compare the population posterior inferred from three simulated "injection" BBH catalogues using the network-output to the results inferred using the analytical model. We create the injection catalogues by sampling events from the distribution characterized by the vector of hyperparameters $\lambda = (-2, 2, 0.5, 0.7)$. We consider the events in each catalogues to be perfectly measured, which means the measured values are the true value with infinite precision. The main difference between the injection sets are the number of events, which are 100, 1000, and 3000, respectively. These numbers are chosen to be approximately the number of detected events one can anticipated in the early-, mid-, and late-2020s [73]. Even though the network can in principle extrapolate to regions outside the trained hyperparameters space, the accuracy is expect to drop, hence the result we obtain will become less credible. Therefore, we choose the range of population prior to be the same as the range where training data are sampled from. We use EMCEE [74] to sample the population posterior [Eq. (6)].

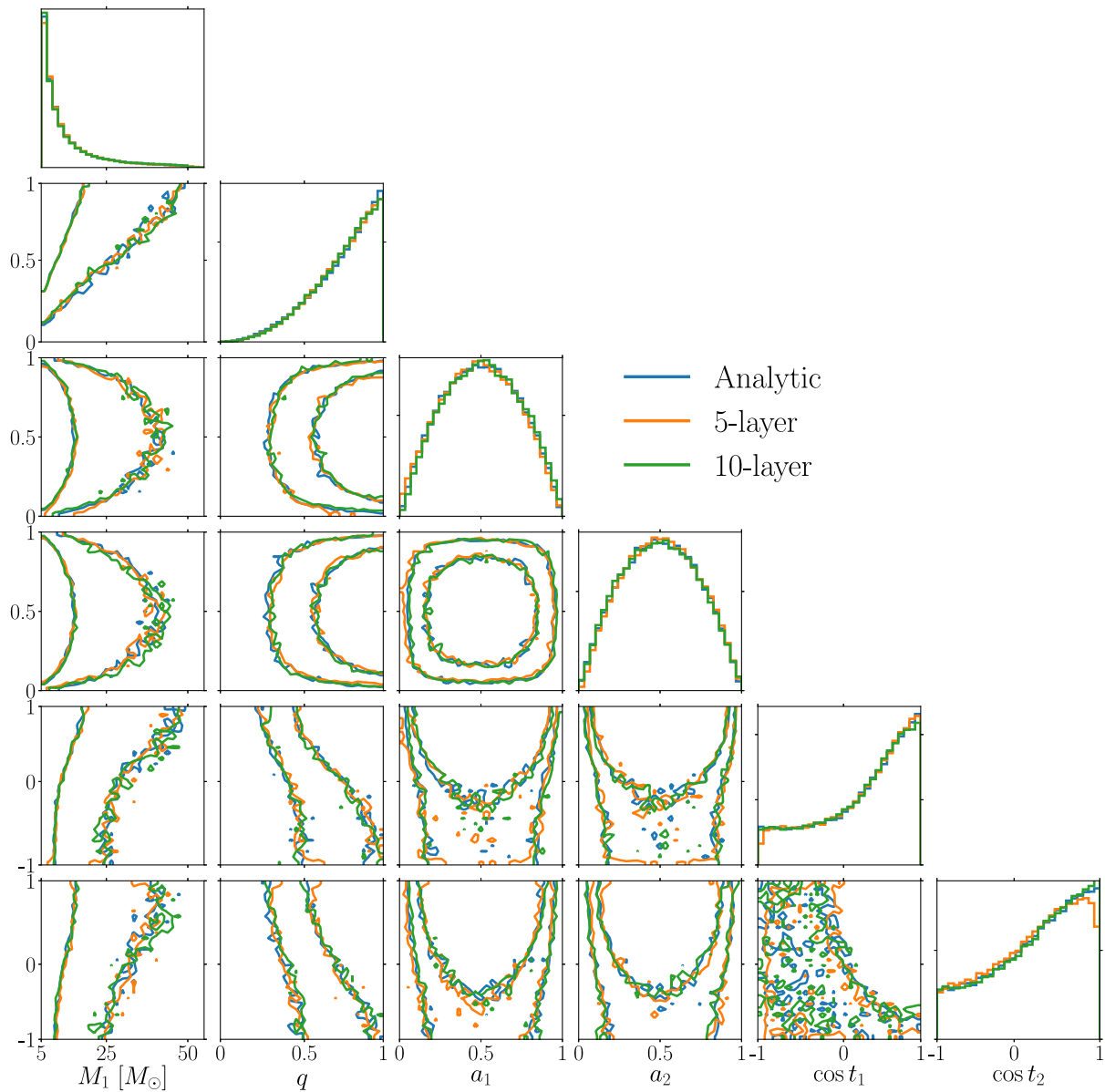

FIG. 2. The joint probability distribution of the 6 event parameters in the phenomenological model, given the hyperparameters $\lambda = (-2, 2, 0.5, 0.7)$. The two contour represent the 68% and 95% confidence interval of the sampled distribution. The analytical (blue) line represents the true answer. The 5-layer (orange) and 10-layer (green) lines are the output of the network with the corresponding number of flow layers. Each contour are created from 100 000 samples, either drawn from the analytical distribution or generated by the network. The 5-layer model fits the analytical answer fairly well on large scale yet having some inaccuracy near the edge of $\cos t_1$ and $\cos t_2$. We find the accuracy of 10-layer model to be sufficient for our purpose.

As shown in Fig. 3, the inference results using a network agree well with the analytical model. Not only the posterior agrees with the injected hyperparameters within the 95% confidence interval, the shape and the location of the confidence interval produced by the network is also similar to the analytical result. There are some minor discrepancies in the case which we consider 3000 events, which can be overcome by increasing the number of layers in the network and training. Considering we are not including other sources of error in this analysis, the minor

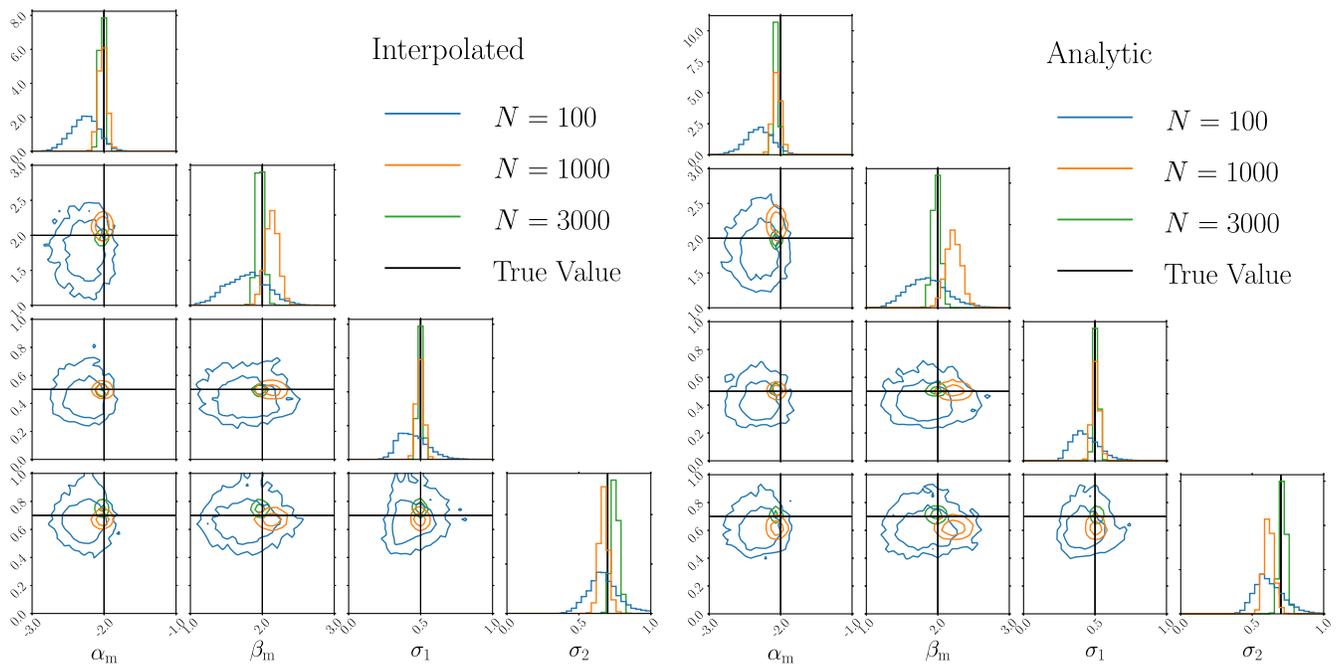

FIG. 3. Left: The population posterior recovered by the neural network emulator as the population likelihood. The injected hyperparameters are $\lambda = (-2, 2, 0.5, 0.7)$, which is marked by the black lines. The contours are the 68% and 95% confidence intervals. The blue, orange, and green line mark the case of having 100, 1000, and 3000 events, respectively. Right: The population posterior recovered by the analytical distribution stated in Sec. IV. The only difference between the right panel and the left panel is the population likelihood, all other factors are the same.

discrepancies shown in Fig. 3 are unlikely to be the dominant source of systematic error.

We have focused on the accuracy aspect of the machine so far. A equally important aspect of this machine is the speed at which a population inference run is done. With 3000 events and each event with 100 posterior samples in a six dimensional event parameters space, one evaluation of the posterior function described in (4) takes ∼0.1 second if we are using the normalizing flow network. On the other hand, we tried using the simulation and the GAUSSIAN_KDE function from SCIPY [75] to evalute the same posterior function, which each evaluation would take more than ∼20 minutes. Note that the simulations used here are relatively simple and fast, which can be generated in a minute. In this case, most of the computational cost comes from estimating the probability of a point in the event parameter-hyper parameter space. A typical inference runs requires ∼$10^3$ to few $10^4$ point to ensure convergence of the results. This means the gaussian kernel density estimation method would require months to years to produce one population inference run, while our network requires only a few hours. More sophisticated simulations will need more time to be produced, which can take days or even weeks on a computer cluster, rendering a direct estimate of $p(\theta|\lambda)$ from simulation impossible.

## V. DISCUSSION

We have incorporated a flow-based deep generative network into a hierarchical Bayesian analysis, and showed that the neural network we have integrated in this study is capable to handle data and simulations which are too complicated for previous machines. While previous studies [24,25] have shown how population-synthesis simulations and machine learning can be used in GW population inference, they are severely limited in various aspects, especially in terms of scalability, thus cannot be used to explore the increasingly complex GW dataset and simulations in practice. We have demonstrated that a normalizing flow network can be used as a highly accurate, efficient, and most importantly scalable machine in GW population inference. There are up to 15 standard event parameters per BBH event in GR, among them there are 8 intrinsic parameters (1 mass and 3 spins per black hole), which are particularly important to astrophysical theories. Many population-synthesis models predict features in at least 4 of these intrinsic parameters, very often in 6 or more intrinsic parameters. Previous approaches can only function accurately and efficiently up to 2 event parameters. When one tries to use the previous machine for 3 or more event parameters, the performance in terms of speed and accuracy drops significantly, and the memory usage starts to become unmanageable. As shown in Sec. IV, our network can reproduce the analytical answer very accurately in a reasonable time even with 6 event parameters. Given the high fidelity of our results and the fact that we will have ∼1000 compact binary coalescence events in late 2020 [73], our method is a promising way forward for simulation-based population inference.

Equipped with this machine, our next step will be testing different state-of-the-art models [76–79] with the upcoming LVC O3 catalogue. An interesting application is that we are now able to compare different family of model directly. Each of the simulations has their own set of hyperparameters. After we obtain the population posterior for each of them, we can marginalize it to obtain the evidence and compare between models. The Bayes factor between one model and another will indicate which model is favored by the data. Furthermore, one can compare these simulation-based models to the evidence from a phenomenological model. In this way we can investigate whether a model contains redundant parameters or not. Being able to compare entire families of models means not only we can use the data to constrain the model, we can also gauge the importance of individual components in a simulation, hence gaining insight on how to improve the simulation.

The simulations used in this work are relatively simple compared to the state-of-the-art models. We also neglected event uncertainties and selection bias, which are crucial when applying this method to real data. The precise effect of the interplay between all these uncertainties and systematics will depend on the properties of the data and simulations. We will follow up with a case study which employs our method and state-of-the-art simulations on O3 data in the soon future, with uncertainties and selection bias taken into account.

We only trained a relatively small network with relatively small amount of data, as compared to other deep neural network trained for the same purpose [41,80]. This means we still have not reached the limit of the network capability. The purpose of this paper is to demonstrate the robustness and efficiency of our method, yet we have not exactly quantified the precise uncertainty from the network as a function of the size of the training data, simulation complexity, and architecture of the network. A study of the precise scaling behavior of the method, which will shed light on the reason why neural networks perform well in this particular type of problem, will be carried out in the future.

This study is specifically dedicated to the GW community, therefore we discussed the BBH case. But the machine can be applied to more general problems such as other GW sources, or even other inference problems that are not related to GWs. As examples, while keeping the vast differences between detector design and potential sources of systematic error in mind, the same machinery can be applied to understanding the population of gamma-ray burst [81], fast radio burst [82], and exoplanets [83], especially when the data is noisy and Bayesian statistics is necessary to characterize the data.

Because of its capability to solve problems with much greater efficiency and accuracy in many settings, deep learning is revolutionizing many different aspects of our society. By integrating the existing tools from deep learning to GW population inference, we enable the possibility of constraining state-of-the-art models with upcoming data at a complexity which was previously intractable. On the other hand, the traditional tools we use to understand physical models and data offer very interpretable checks, which we can use to gauge the performance of our deep learning model and improve it. GW data analysis is known to be in the extremely noise dominated regime, which is where deep learning often encounter trouble in. We hope that the progressively frequent cross talk between the deep learning and GW communities can be mutually beneficial to both fields.

## ACKNOWLEDGMENTS

The authors thanks V. Baibhav, T. Helfer, E. Berti, D. Gerosa, K. K. Y. Ng, P. T. H. Pang, C. Creque-Sarbinowski, Will Farr, David Spergel for constructive feedbacks. K. W. K. Wong is supported by NSF Grants No. PHY-1912550 and AST-1841358, NASA ATP Grants No. 17-ATP17-0225 and No. 19-ATP19-0051, and NSF-XSEDE Grant No. PHY-090003. S. H. and G. C. are supported by the Simon Foundations. Computational work was performed at the Maryland Advanced Research Computing Center (MARCC). This work has received funding from the European Union's Horizon 2020 research and innovation programme under the Marie Skłodowska-Curie Grant Agreement No. 690904. The authors would like to acknowledge networking support by the GWverse COST Action CA16104, "Black holes, gravitational waves and fundamental physics".